# Period doubling of multiple dissipative-soliton-resonance pulses in a fiber laser

**LIMING HUA,**[1] **SHUAI WANG,**[1] **XU YANG,**[1] **XIONGQUAN YAO,**[1] **LEI LI,**[1,*] **ANDREY KOMAROV,**[2] **MARIUSZ KLIMCZAK,**[3,4] **DEYUAN SHEN,**[1] **DINGYUAN TANG,**[5] **LEI SU,**[6] **AND LUMING ZHAO**[1,6,*]

[1]*Jiangsu Key Laboratory of Advanced Laser Materials and Devices, Jiangsu Collaborative Innovation Center of Advanced Laser Technology and Emerging Industry, School of Physics and Electronic Engineering, Jiangsu Normal University, Xuzhou, 221116 Jiangsu, P. R. China*
[2]*Institute of Automation and Electrometry, Russian Academy of Sciences, Academician Koptyug Prospekt 1, 630090 Novosibirsk, Russia*
[3]*Faculty of Physics, University of Warsaw, Warsaw, Poland*
[4]*Institute of Electronic Materials Technology, Wólczynska 133, 01-919 Warsaw, Poland*
[5]*School of Electrical and Electronic Engineering, Nanyang Technological University, Singapore 639798*
[6]*School of Engineering and Materials Science, Queen Mary University of London, London E1 4NS, UK*
*\* sdulilei@gmail.com, lmzhao@ieee.org*

**Abstract:** We report on the experimental observation of period doubling of multiple dissipative-soliton-resonance (DSR) pulses in an all-normal-dispersion fiber laser based on a nonlinear amplifying loop mirror. By increasing the pump power, in addition to the typically linearly pulse broadening under fixed pulse peak power, the jump from a single DSR pulse to multiple DSR pulses was observed. During this process, the period doubling appears, that is, the DSR pulses can exhibit the characteristics of period doubling in a multi-pulse state. The typical DSR performance of linearly pulse duration change versus pump power varying can be maintained when the period doubling of multiple DSR pulses appears.

## 1. Introduction

Ultrafast fiber lasers as an attractive ultrashort pulse source have attracted extensive attention due to significant applications in areas of micro-machining [1], biomedicine [2, 3] etc. Ultrafast fiber lasers not only have important applications, but also are an excellent platform for investigation on nonlinear optical systems. Ultrafast fiber lasers can be used to observe a variety of nonlinear phenomena, such as period doubling bifurcation. The period doubling bifurcation is a typical route to chaos, and it is a typical stable state in nonlinear systems. The period doubling in fiber lasers was first reported by Tamura et al [4]. By introducing a bandpass filter in a negative dispersion fiber laser, Tamura et al. realized the period doubling of solitons. The period doubling bifurcation can be generated in fiber lasers operated in different dispersion regions and is independent of the mode-locking mechanism and wavelength. Using the nonlinear polarization rotation (NPR) technique, Zhao et al. observed the period doubling bifurcation in a dispersion-managed soliton fiber laser for the first time [5]. In addition, Zhao et al. also reported period doubling bifurcation of bound states of solitons, period doubling bifurcation of multiple solitons, and period doubling bifurcation of gain-guided solitons, all of which are studied in Erbium-doped mode-locked fiber lasers based on the NPR technique [6-8]. Wu et al. observed period doubling bifurcation of vector solitons in a thulium-doped fiber laser using carbon nanotube mode-locking [9].

The excessively accumulated nonlinear effects in ultrafast fiber lasers will restrict the increasing of pulse energy, which leads to pulse splitting. In 2008, Chang et al. theoretically discovered dissipative-soliton-resonance (DSR) area where pulse energy can be increased indefinitely without pulse splitting [10]. In the DSR regime, the peak of the pulse remains constant while the pulse width broadens linearly with the increase in the pump power [11-13]. In fact, under certain conditions, DSR will undergo pulse splitting into a multi-pulse state [14, 15]. Komarov et al. theoretically investigated multi-pulse state of DSR harmonic mode-locking, and they concluded that the number of DSR pulses in the steady state is related to the initial condition, regardless of the increasing pump power [14]. Wang et al. experimentally demonstrated the multiple-pulse DSR evolving from the original DSR pulse in an all-normal-dispersion fiber laser [15].

Period doubling of DSR pulses was recently reported by Wang et al [16]. Therefore, it is interesting to know whether the period doubling could be maintained while multiple DSR pulses appears. In this paper we further report on the experimental observation of period doubling of multiple DSR pulses in an all-normal-dispersion fiber laser. The fiber laser is mode-locked by a nonlinear amplifying loop mirror (NALM). Since the generation of DSR in a fiber laser is the result of strong peak power clamping effect [17], its peak power is relatively low, and the accumulated nonlinear effects are not significant. Therefore, 380 m optical fiber are inserted in the cavity to increase the accumulated nonlinear phase shift. When the nonlinear phase shift is large enough, which leads to the considerable nonlinear effects, multi-pulse DSR and period doubling of multi-pulse DSR can be observed. To the best of our knowledge, so far no report on period doubling of multi-pulse DSR was presented.

## 2. Experimental setup

Figure 1 shows the schematic of the all-normal-dispersion fiber laser used in our experiment.

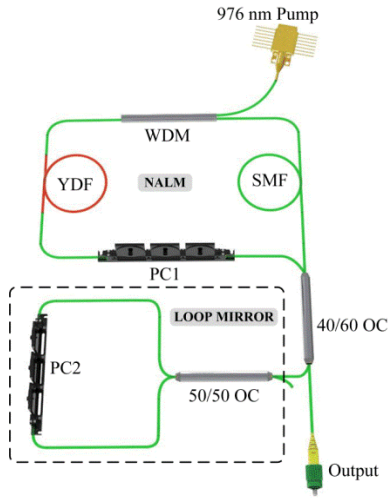

Fig. 1. Schematic of the all-normal-dispersion fiber laser. WDM: wavelength division multiplexer; YDF: ytterbium-doped fiber; PC: polarization controller; OC: fiber coupler; SMF: 380-m-long single-mode fiber.

A NALM (in the top of Fig. 1), which acts as a fast saturable absorber, is coupled into a loop mirror (in the dashed line frame) through a 40/60 fiber coupler (OC). In the NALM, the gain is provided by using a 45-cm-long single-cladding, ytterbium-doped fiber (YDF, CorActive Yb501) with a core absorption 139 dB/m at 915 nm. The YDF is pumped through a wavelength division multiplexer (WDM) using a 976 nm pigtailed laser diode, providing a maximum pump power of 600 mW. A 380-m-long single-mode fiber (SMF, Nufern, 1060-XP) inserted in the NALM is used to enhance the nonlinear effects. A fiber coupler with 40% port acting as an output is placed after PC1.The loop mirror which serves as a reflective mirror comprises a polarization controller (PC2) and a 50/50 OC. The fiber pigtail of all the optical components in this laser is HI1060. The total cavity length is ∼405 m, corresponding to a fundamental repetition rate of 494 kHz. The net cavity dispersion is calculated to be 8.86 ps$^2$. A 1 GHz oscilloscope (Agilent, DSO9104H), a 1.2GHz photo-detector (Thorlabs, DET01CFC/M), an optical spectrum analyzer (OSA, Yokogawa AQ6317C), an optical power meter (Thorlabs, PM100D), and a radio frequency (RF) signal analyzer (Agilent, N9320B) are used to monitor the output of the mode-locked pulse trains.

## 3. Experimental results and discussion

In our experiment, single DSR pulse can be easily obtained because of large normal dispersion. Provided that the orientations of the polarization controllers are appropriately set, period doubling of single DSR pulse can be obtained by simply increasing the pump power beyond the mode-locking threshold [16]. Figure 2 shows as example an experimentally observed period doubling of single DSR pulse. Figure 2(a) shows the oscilloscope trace of the period doubling of single DSR pulse at a pump power of 28.5 mW. The details of the single DSR pulse are shown in Fig. 2(b) and Fig. 2(c), which corresponds to the two distinct states in Fig. 2(a) (indicated by arrow). The pulse temporal profile exhibits an asymmetrical structure, as the pulse front has a higher intensity than the pulse trailing edge. Figure 2(d) shows the corresponding optical spectrum of the period doubling of single DSR pulse. The inset in Fig. 2(d) shows the typical bell-shaped optical spectrum of the DSR pulse [18-22]. The central wavelength is 1045.5 nm and the 3-dB bandwidth of the spectrum is about 0.039 nm. The signal near 1095 nm is due to the stimulated Raman scattering (SRS) effect. The long-cavity configuration can lead to the low threshold required for the SRS. Figure 2(e) shows the RF spectrum of the laser output. In addition to the fundamental repetition rate, a new frequency component appears at half of the fundamental repetition frequency and its harmonic, showing that the single DSR pulse is in the state of period doubling.

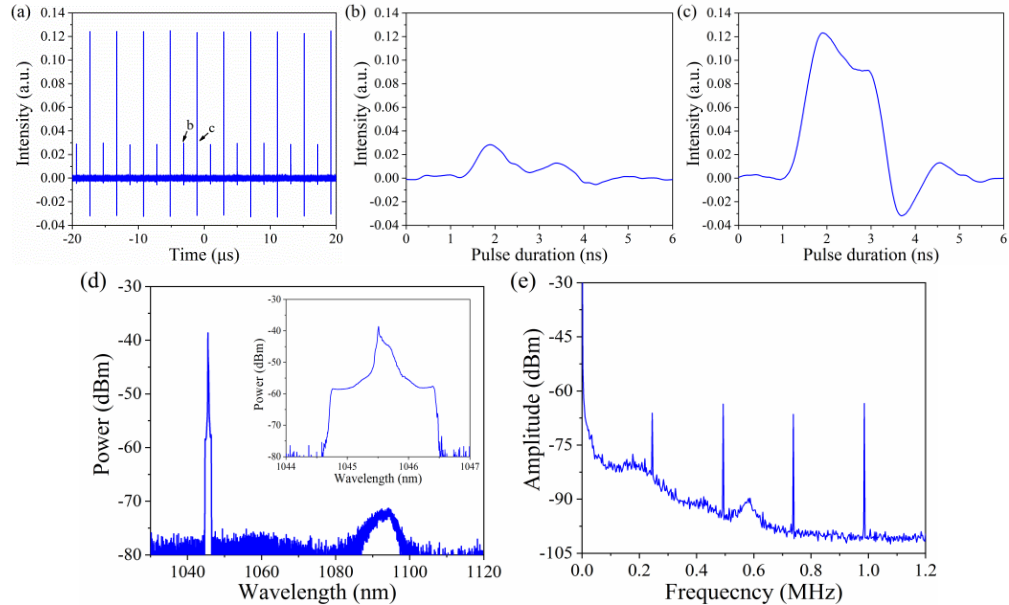

Fig. 2. Period doubling of single DSR pulse. (a) Pulse train; (b)(c) Corresponding to the details of single DSR pulse under two different states; (d) Corresponding optical spectrum. Inset: local amplification of the corresponding optical spectrum; (e) RF spectrum with a span of 1.2 MHz and a resolution bandwidth of 10 Hz.

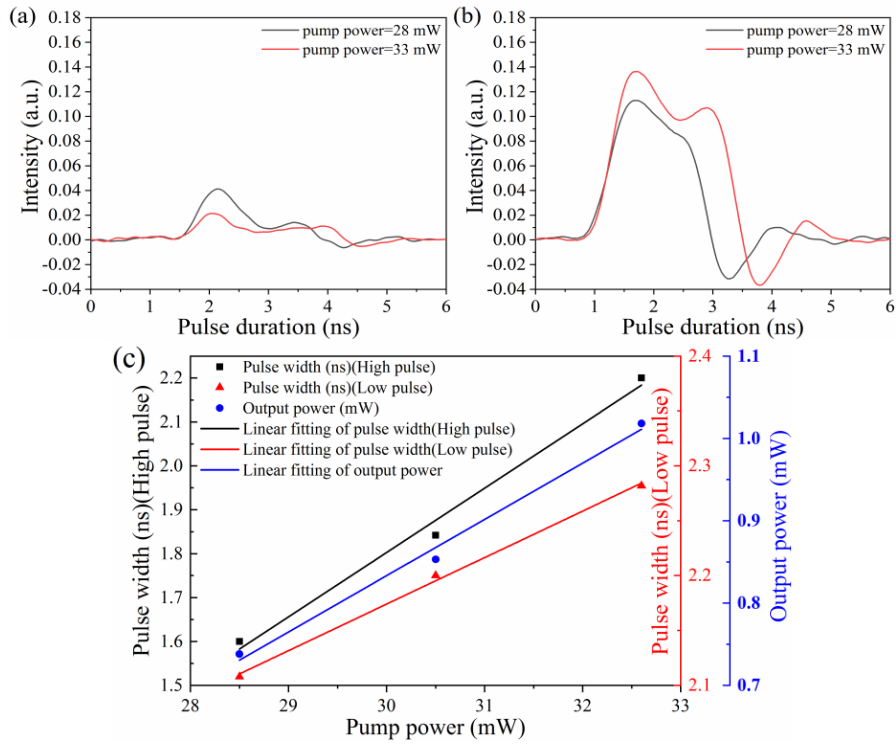

Fig. 3. Tuning the pump power from 28 mW to 33 mW. (a)(b) Oscilloscope trace of pulse evolution under two different states; (c) Pulse width and average output power versus pump power.

With PC paddles being fixed, the single DSR pulse broadens gradually with the increase in the pump power from 28 to 33 mW. In this process, the period doubling always exists, as shown in Fig. 3. Low-intensity pulse broadens from 2.1 ns to 2.3 ns (Fig. 3(a)), while high-intensity pulse broadens from 1.6 ns to 2.2 ns (Fig. 3(b)). In Fig. 3(c) pulse width and the average output power are plotted in function of the pump power delivered to the active fiber. Both the pulse width and the average output power are proportional to the pump power, and can be approximated with a linear fitting.

As pump power continues to increase, single DSR pulse jumps into dual-pulse DSR, and eventually becomes period doubling of dual-pulse DSR. The period doubling of dual-pulse DSR is depicted in Fig. 4. Figure 4(a) shows the oscilloscope trace of the period doubling of dual-pulse DSR at a pump power of 44 mW. There are two DSR pulses existed in the cavity with pulse separation of about 20 ns. The details of dual-pulse DSR are shown in Fig. 4(b) and Fig. 4(c), which corresponds to the two distinct states in Fig. 4(a) (indicated by arrow). The corresponding optical spectrum and RF spectrum of the period doubling of dual-pulse DSR are shown in Fig. 4(d) and Fig. 4(e). The central wavelength is slightly red-shifted to 1045.8 nm and the 3-dB bandwidth of the spectra is broadened to about 0.061 nm. The RF spectrum confirms that the dual-pulse DSR is in period doubling state.

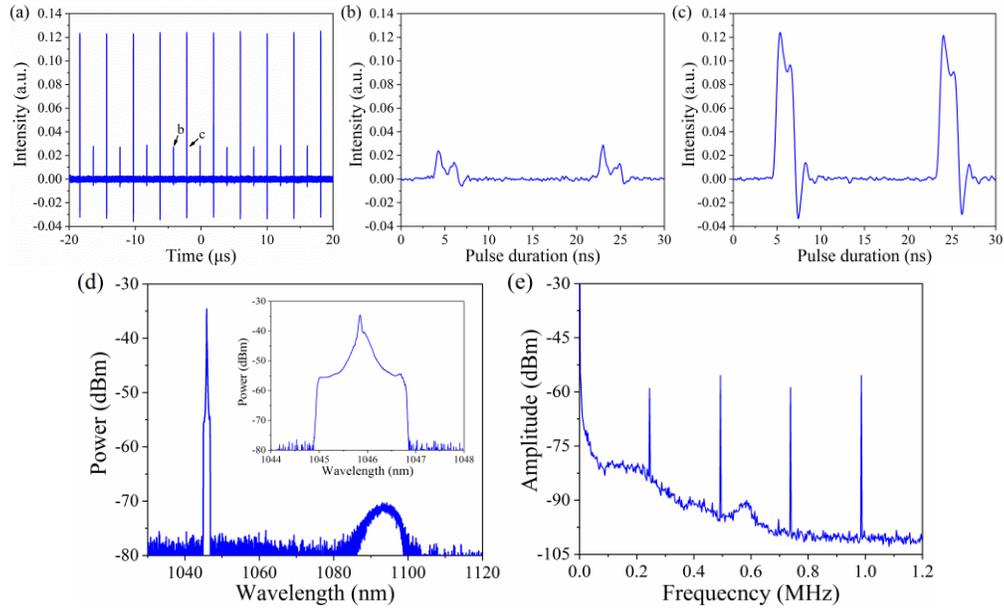

Fig. 4. Period doubling of dual-pulse DSR. (a) Pulse train; (b)(c) Corresponding to the details of dual-pulse DSR under two different states; (d) Corresponding optical spectrum. Inset: zoom-in of the corresponding optical spectrum; (e) RF spectrum with a span of 1.2 MHz and a resolution bandwidth of 10 Hz.

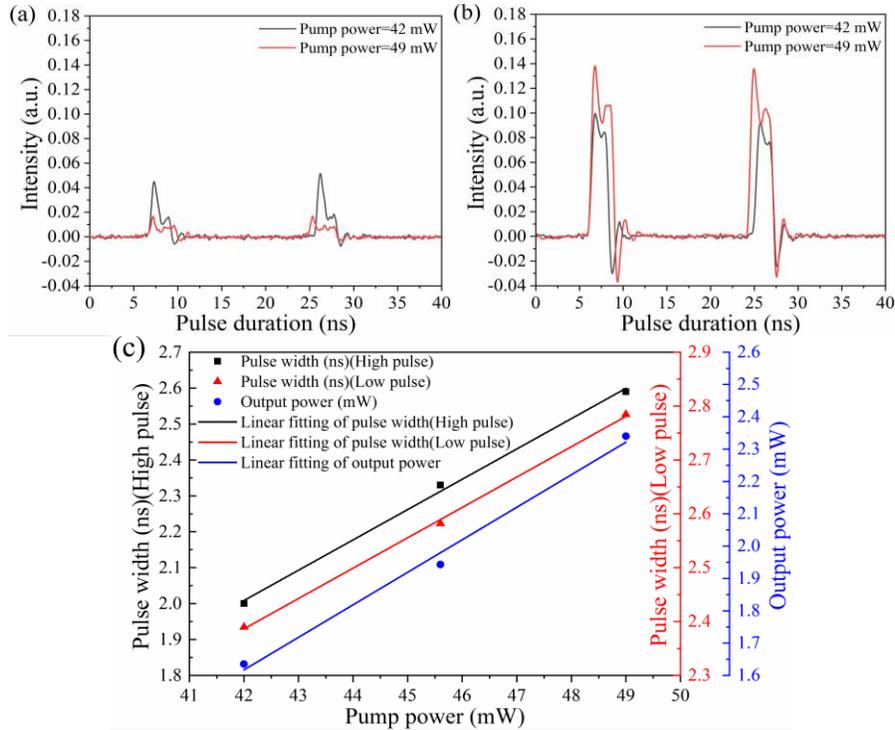

Fig. 5. Tuning the pump power from 42 mW to 49 mW. (a)(b) Oscilloscope trace of pulse evolution under two different states; (c) The pulse width of the first pulse of the low-intensity pulses in the state of dual-pulse, the pulse width of the first pulse of the high-intensity pulses in the state of dual-pulse, and average output power in function of pump power.

It is found that similar to the case of period doubling of single DSR pulse, the dual-pulse DSR also can broaden linearly with the increase in the pump power while maintaining the period doubling state. As shown in Fig. 5(a) and Fig. 5(b), when pump power increased from 42 mW to 49 mW, the first pulse of low-intensity pulses in the state of dual-pulse broadens from 2.4 ns to 2.8 ns, while the first pulse of high-intensity pulses in the state of dual-pulse broadens from 2 ns to 2.6 ns. The pulse width and average output power increase linearly with respect to the pump power, and can be approximated with a linear fitting (Fig. 5(c)). We note that the pulse separation between the two DSR pulses is reduced during the pump power increasing, while the pulse intensity difference between the low-intensity pulses and the high-intensity ones increased.

Further increasing the pump power, three DSR pulses could be obtained. The period doubling of three DSR pulses could also be achieved. Figure 6 depicts the case of period doubling of three DSR pulses at a pump power of 57 mW. Figure 6(a) shows the oscilloscope trace where there are three DSR pulses coexisting in the cavity. Details of the three DSR pulses are shown in Fig. 6(b)-(c). The optical spectrum is shown in Fig. 6(d). The central wavelength is further red-shifted to 1047.8 nm but the 3-dB bandwidth of the spectra reduced to about 0.050 nm. The RF spectrum again confirms that the appearance of period doubling (Fig. 6(e)).

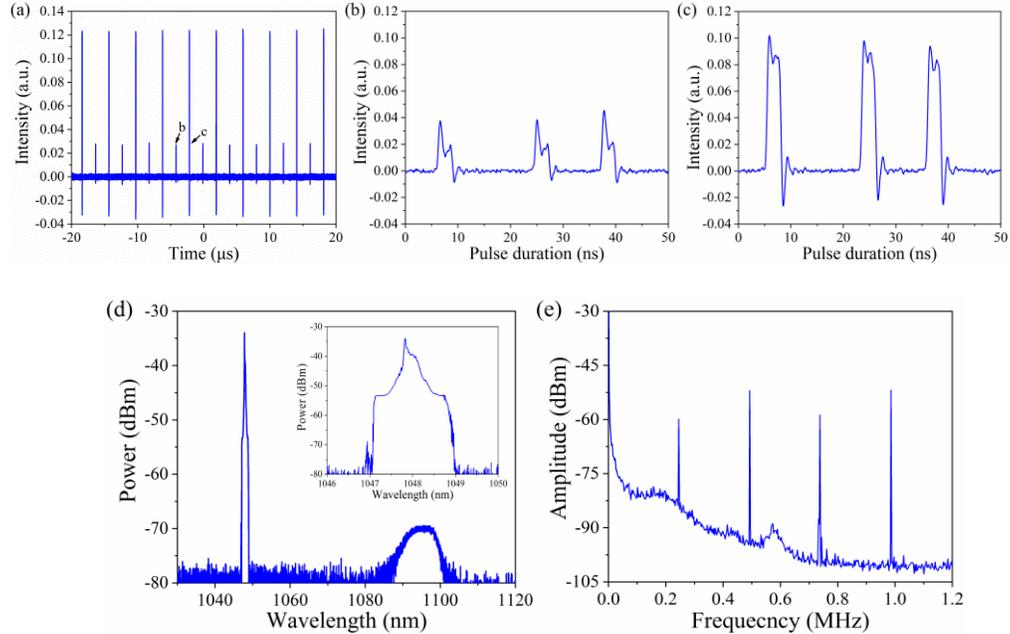

Fig. 6. Period doubling bifurcation of three-pulses DSR. (a) Pulse train; (b)(c) Corresponding to the details of three-pulses DSR under two different states; (d) Corresponding optical spectrum. Inset: local amplification of the corresponding optical spectrum; (e) RF spectrum with a span of 1.2 MHz and a resolution bandwidth of 10 Hz.

Tuning the pump power from 57 mW to 61mW, the three pulses can maintain period doubling with gradually broadening of pulse width. We measured the first pulse in the state of three pulses, the low-intensity pulses broaden from 2.7 ns to 3 ns, while high-intensity one broadens from 2.3 ns to 2.6 ns (Fig. 7(a)-(b)). The pulse width and average output power increase linearly with respect to the pump power as shown in Fig. 7(c), which can be approximated with a linear fitting. We note that the pulse separation is reduced when the pump power increases.

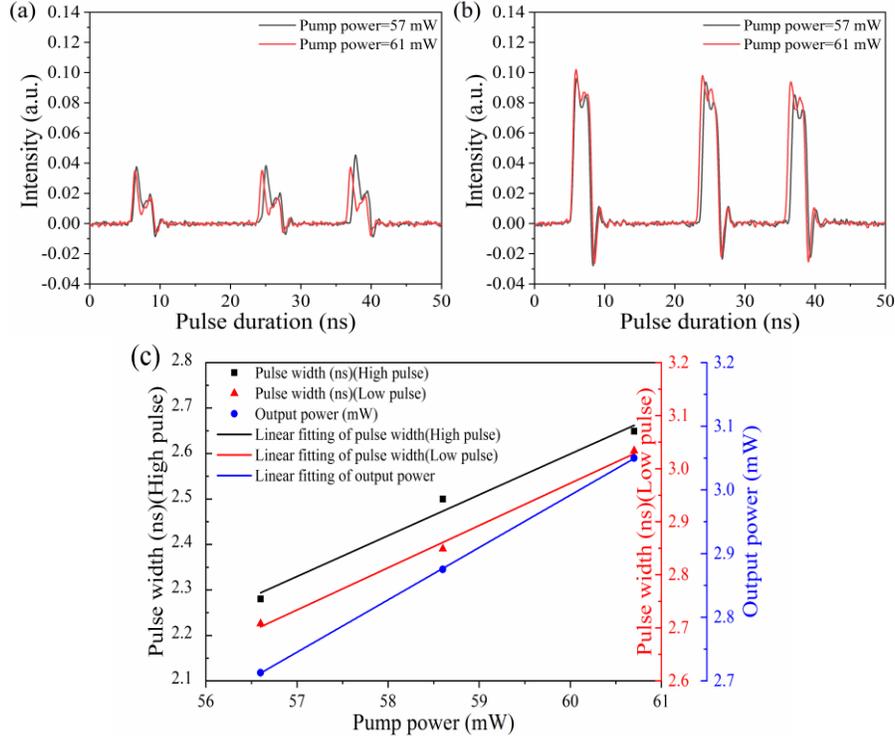

Fig. 7. Tuning the pump power from 57 mW to 61 mW. (a)(b) Oscilloscope trace of pulse evolution under two different states; (c) The pulse width of the first pulse of the low-intensity pulses in the state of three pulses, the pulse width of the first pulse of the high-intensity pulses in the state of three pulses, and average output power in function of pump power.

We note that the development from period doubling of single DSR pulse to period doubling of double DSR pulses, and then to period doubling of three DSR pulses can always be repeated even after the mode locking state was destroyed due to over pumping. However, the achievement of period doubling of single DSR pulse is the precondition for obtaining such a route. If the original single DSR pulse does not exhibit period doubling, no period doubling of multiple DSR pulses can be obtained with increasing pump power.

## 4. Conclusion

We experimentally observed period doubling of multi-pulse DSR in an all-normal-dispersion fiber laser based on a NALM. It is found that by increasing the pump power, in addition to the typical pulse broadening at the fixed pulse peak power, a single DSR pulse can evolve into a multi-pulse state. In this process, the DSR pulses can exhibit period doubling in a multi-pulse state. So far period doubling of dual-pulse and three-pulses can be observed. Further increasing the pump power will destroy the mode locking state. When the period doubling of multiple pulses are achieved, slightly increasing the pump power will make the pulse presenting linearly pulse duration change versus the pump power. In another word, the DSR performance could be maintained when the multiple pulses exhibits period doubling behavior. The experimental observations great enrich the dynamics of DSR pulses in fiber lasers.

## Funding

Protocol of the 37th Session of China-Poland Scientific and Technological Cooperation Committee (37-17); European Union's Horizon 2020 research and innovation programme under the Marie Skłodowska-Curie grant agreement No. 790666; Priority Academic Program Development of Jiangsu Higher Education Institutions (PAPD); Postgraduate research innovation program of Jiangsu Normal University (2018YXJ592). Lei Li and Luming Zhao acknowledge support from Jiangsu Overseas Visiting Scholar Program for University Prominent Young & Middle-aged Teachers and Presidents. Mariusz Klimczak acknowledges support from Fundacja na rzecz Nauki Polskiej (FNP) in scope of First TEAM/2016-1/1 (POIR.04.04.00-00-1D64/16).

**Disclosures**

The authors declare that there are no conflicts of interest related to this article.